\documentstyle[epsf,prb,twocolumn,aps]{revtex}

\relax
\newcommand{\be}{\begin{equation}}
\newcommand{\ee}{\end{equation}}
\newcommand{\bea}{\begin{eqnarray}}
\newcommand{\eea}{\end{eqnarray}}

\newcommand{\re}{{\rm e}}

\begin{document}
\twocolumn[\hsize\textwidth\columnwidth\hsize\csname @twocolumnfalse\endcsname
\title{Realization of Haldane's Exclusion
Statistics in a Model of Electron-Phonon
Interactions}
\author{Catherine P\'epin and Alexei M. Tsvelik}
\address{Department of Physics, University of Oxford, 1 Keble Road, Oxford, \\
OX1 3NP, UK}
\date{\today }
\maketitle

\begin{abstract}
We discuss an integrable model describing one-dimensional electrons
interacting with two-dimensional anharmonic phonons. In the low temperature
limit it is possible to decouple  phonons and consider 
one-dimensional excitations separately. 
They  have  a trivial two-body scattering
matrix and obey fractional statistics. As far as we know the original 
 model presents the
first example of a model with local bare interactions generating purely
statistical interactions between renormalized particles. As a by-product we
obtain non-trivial thermodynamic equations for the interacting system of
two-dimensional phonons.
\end{abstract}
\newpage
\vskip -0.1truein]
The concept of fractional exclusion statistics
was introduced by Haldane in 1991\cite{Haldane}
and since then there has been a continuous
interest in this subject (see, for example
\cite{Ha,Schoutens,Frahm}). In
the ideal case of a `quantum gas of
non-interacting particles satisfying fractional
exclusion statistics' (the Abelian one) the
number of states $ Q_{a}$ of particles of
species $a$ linearly depends on numbers of
occupied states of particles of species $b$,
$N_{b}$ with the same energy: 
\[
\partial Q_{a}/\partial N_{b}=-G_{ab}
\]
where $G_{ab}$ is a matrix of statistical interactions. As has been
demonstrated by many authors, \cite
{Wu} the thermodynamics of such an ideal gas of anyons is described by the
set of equations 
\bea
\epsilon _{a}(E)+G_{ab}T\ln [1+{\rm e}^{-\epsilon _{b}(E)/T}]=E \label{excl}
\eea
\[
F/L=-T\sum_{a,E}\ln [1+{\rm e}^{-\epsilon _{a}(E)/T}]
\]

 It has been emphasised by different authors, 
most  recently by Frahm \cite{Frahm}
and Bouwknegt and 
Schoutens \cite{Schoutens}
that integrable systems in (1+1)-dimensions represent natural
realizations of all sorts of exotic Abelian and even non-Abelian
exclusion statistics (the latter ones are not discussed in this
paper). 
This follows from the fact that creation 
and annihilation operators
of renormalized  particles  in these  systems
 satisfy commutation relations which explicitly depend on  the
two-particle $S$-matrix (the Zamolodchikov-Faddeev algebra) being thus
neither bosonic nor fermionic: 
\begin{eqnarray}
Z^+_a(\theta_1)Z^+_b(\theta_2) &=& S^{\bar a,\bar b}_{a,b}(\theta_1 -
\theta_2)Z^+_{\bar b}(\theta_2)Z^+_{\bar a}(\theta_1)\label{faddeev}\\
Z^a(\theta_1)Z^b(\theta_2) &=& S_{\bar a,\bar b}^{a,b}(\theta_1 -
\theta_2)Z^{\bar b}(\theta_2)Z^{\bar a}(\theta_1)\nonumber\\
Z^a(\theta_1)Z^+_b(\theta_2) &=& S^{\bar b, a}_{b, \bar a}(\theta_2 -
\theta_1)Z^+_{\bar b}(\theta_2)Z^{\bar a}(\theta_1) + \delta(\theta_1
- \theta_2)\delta_{a}^{b} \nonumber
\end{eqnarray}
Here the operators 
 $Z_a^+(\theta)$ and $Z_a(\theta)$ create (annihilate) the state
with rapidity $\theta$ and isotopic number $a$ (recall that rapidity 
is a relativistic  parametrization of  energy and momentum  $E =
 m\cosh\theta, P = m\sinh\theta$; for simplicity we shall discuss only
 Lorentz invariant models, but the results have more general
 validity).

Unfortunately, most known integrable models are not good pedagogical
examples of the effects of exclusion statistics because they describe {\sl %
interacting} anyons , where the interaction derives from a momentum
dependent   $S$-matrix. This is why the free energy in these theories is
described by a complex system of nonlinear integral equations which in
general, cannot be solved in closed form. The $S$-matrix in this
models becomes purely elastic only  at high momentum: 
\begin{equation}
S_{b,a}^{\bar{b},\bar{a}}(\theta )\rightarrow \delta _{a}^{\bar{a}}\delta
_{b}^{\bar{b}}\exp [-{\rm i}\pi sign\theta
G_{ab}],\;\,\,\,\,\,\,\,\,\,(|\theta |\rightarrow \infty ).  \label{elastic}
\end{equation}
 A model where this  helds at all rapidities, 
{\sl would} be an ideal gas of anyons. Examples of such models known so far
involve long-range interactions (for instance, the Haldane-Shastry
spin-chain \cite{Shastry}; see also the discussion in \cite{Ha}).

In this paper we provide an example of an integrable model with {\it local
interactions} where the two-body $S$-matrix of physical particles has the
form (\ref{elastic}) from the very start. We study a limiting case of an
integrable  model first  suggested by Fateev \cite{fateev1} . The model
describes a one -dimensional sea of interacting fermions, coupled to a {\sl %
\ two }dimensional bath of anharmonic phonons. The Hamiltonian density has
the following form: 
\begin{equation}
{\cal H}={\cal H}_{e}+{\cal H}_{p}+{\cal H}_{I},  \label{model1}
\end{equation}
where
\[
{\cal H}_{e}=
\]
\[
-{\rm i}R^{+}\partial _{x}R+{\rm i}L^{+}\partial _{x}L+\left(
2g/\pi \right) R^{\dagger }RL^{\dagger }L - \mu(R^+R + L^+L)
\]
describes a one dimensional electron gas of right ($R$), and left-moving ($L$%
) fermions (``Thirring model'') with the chemical potential $\mu$, and 
\begin{eqnarray}
{\cal H}_{p}= &&\frac{1}{2}\sum_{n=1}^{N}\{P_{n}^{2}+(\partial _{x}\phi
_{n})^{2}\}+(m/\beta )^{2}\sum_{n=1}^{N-1}{\rm e}^{\beta (\phi _{n+1}-\phi
_{n})}  \nonumber \\
&&+\frac{1}{2}(m/\beta )^{2}[{\rm e}^{2\beta \phi _{1}}+{\rm e}^{-2\beta
\phi _{N}}],
\end{eqnarray}
where $[P_{n}(x),\phi _{m}(y)]=-{\rm i}\delta _{nm}\delta (x-y),$ describes $%
N$ parallel chains coupled to each other by an anharmonic nearest-neighbour
interaction , (``Toda lattice''), and

\[
{\cal H}_{I}=m(R^{\dagger }L+\;L^{\dagger }R){\rm e}^{\beta \phi _{1}}
\]
describes the coupling between the two-dimensional phonon gas and the Fermi
sea. The pure fermionic Hamiltonian corresponds to the massive Thirring
model which is one of the first integrable models known and has numerous
applications in condensed matter physics. \ss 

 In the limit $\beta << 1$ one can expand in $\beta$. In the limit 
 $ N \rightarrow \infty$ the phonon spectrum becomes continuous 
 two-dimensional  with the
 dispersion law given by 
\bea
\omega^2(p)
 = p_x^2 + m^2\sin^2(\pi p_y/2) \label{disp}
\eea
 As we shall see this form of dispersion 
 remains robust at any $\beta$.  We shall be interested in the case
 when $\epsilon_F = \mu - M > 0$ when  there are charged particles in
 the ground state. It turns out that in this case  phonons decouple
 and one can study the remaining edge excitations
 separately. In terms of the original fermions the corresponding 
 effective action describes  one-dimensional fermions interacting by a
 phonon-mediated interaction  non-local both in time and space. For
 instance, in the
 leading order in $\beta$ we have 
\bea
<<\re^{\beta\phi_1(\tau,x)}\re^{\beta\phi_1(0,0)}>> =
 - \frac{\beta^2}{4\pi(\tau^2 + x^2)^{3/2}}
\eea
The exact solution  gives the most illuminating
 results for thermodynamics demonstrating that true edge excitations 
are ideal anyons. Namely, there are 
two oppositely charged particles realizing Haldane's 
exclusion statistics. 
Particles with the same charge scatter with phase shift
$g$, whilst particles with opposite charge scatter with phase
shift $\pi- g$, such that 
\begin{eqnarray}
G = \left( 
\begin{array}{cc}
g/\pi & 1 - g/\pi \\ 
1 - g/\pi & g/\pi
\end{array}
\right)  \label{G}
\end{eqnarray}

 We would like to mention here an amusing feature of model
 (\ref{model1}): under the transformation 
\begin{equation}
g\rightarrow \pi -g,(\beta \rightarrow 4\pi /\beta ) \label{dual}
\end{equation}
the fermionic part of model (\ref{model1}) transforms into a nonlinear sigma
model of a single complex bosonic field and the phonon part is transformed
into a Toda chain with $N-1$ sites (one site shorter) so that the
corresponding Lagrangian density becomes, according to Fateev \cite{fateev1}%
, 
\begin{eqnarray}
{\cal L} &=&\frac{1}{2}\frac{\partial _{t}\chi ^{*}\partial _{t}\chi -%
\partial _{x}\chi ^{*}\partial _{x}\chi }{1+(\beta /2)^{2}\chi ^{*}\chi }-%
\frac{m^{2}}{2}|\chi |^{2}{\rm e}^{\beta \phi _{1}}  \nonumber \\
&&+\frac{1}{2}\sum_{n=1}^{N-1}\{(\partial _{t}\phi _{n})^{2}-(\partial %
_{x}\phi _{n})^{2}\}-(m/\beta )^{2}\sum_{n=1}^{N-2}{\rm e}^{\beta (\phi
_{n+1}-\phi _{n})}  \nonumber \\
&&+\frac{1}{2}(m/\beta )^{2}[{\rm e}^{\beta \phi _{1}}+{\rm e}^{-\beta \phi
_{N-1}}]  \label{model2}
\end{eqnarray}
Therefore a weak coupling limit of the fermionic model is dual to the strong
coupling limit of the bosonic model (\ref{model2}) with $N-1$ sites and vice
versa. This duality is reproduced in  our results serving  as a
check for correctness of  the exact solution. 
The duality was obtained by Fateev in essentially nonperturbative way
and it remains unclear at the moment how to relate the fields of model (\ref
{model2}) to the original fermions and bosons.

Let us recall the relevant results from (\cite{fateev1}). The spectrum
consists of a charged particle (with the corresponding anti-particle) of
mass $M$ and $N-1$ neutral particles with masses 
\begin{equation}
M_{a}=2M\sin (\pi a/2N),a=1,2,...N-1.
\end{equation}
where $M$ is related to the bare mass $m$. 
The latter excitations  are particle-hole bound states. It is very easy to
establish a connection between these particles and the phonons of the Toda
lattice: the above spectrum is reproduced in the limit $\beta \rightarrow 0$
when the phonons become harmonic (see Eq.(\ref{disp}). 
It is a well known fact in the theory of
Toda models that the exact spectrum coincides with the one obtained in the
harmonic approximation (see, for example, \cite{arinstein}, \cite{braden}, 
\cite{sasaki}; see also \cite{klassen} for general discussion of
thermodynamics of theories with elastic $S$-matrices). The scattering is
purely elastic (all $S$ matrices are diagonal in isotopic indices).

The Thermodynamic Bethe Ansatz (TBA) equations for models with elastic
scattering has the following standard form (see \cite{klassen} for a
review). The free energy of the system of length $L$ is given by 
\begin{eqnarray}
F/L = - T\sum_a \frac{m_a}{2\pi}\int {\rm d}\theta \cosh\theta \ln[1 + {\rm e%
}^{- \epsilon_a(\theta)/T}]
\end{eqnarray}
where the excitation energies are determined from the following nonlinear
integral equations: 
\begin{eqnarray}
T\ln[1 + {\rm e}^{\epsilon_n(\theta)/T}] - (\alpha\delta_{nm} + G_{nm})*T\ln[%
1 + {\rm e}^{-\epsilon_m(\theta)/T}]  \nonumber \\
= m_n\cosh\theta  \label{TBA}
\end{eqnarray}
\[
G_{nm}(\theta) = - \frac{1}{2\pi{\rm i}}\frac{\partial\ln S_{nm}(\theta)}{%
\partial\theta} 
\]
where $\alpha = 1$ for fermions and $\alpha = 0$ for bosons (that is for the
neutral bound states) and the star denotes convolution: 
\[
G*f(\theta) \equiv \int_{-\infty}^{\infty}G(\theta -
\theta^{\prime})f(\theta^{\prime}){\rm d}\theta^{\prime}
\]

From Ref.\cite{fateev1} we obtain the following expressions for the Fourier
transformations of the kernels: 
\[
G_{ab}(\omega) = \frac{4\sinh(g\omega/2N)\sinh[(\pi - g)\omega/2N]}{%
\sinh(\pi\omega/2N)} 
\]
\[
\times\frac{\cosh[(N - max(a,b))\pi\omega/2N]\sinh[(min(a,b)\pi\omega/2N]}{%
\cosh (\pi\omega/2)} 
\]
\[
G_{a,+}(\omega) = G_{a,-}(\omega) = 
\]
\[
\frac{2\sinh(g\omega/2N)\sinh[(\pi - g)\omega/2N]}{\sinh(\pi\omega/2N)}\frac{%
\sinh (a\pi\omega/2N)}{\cosh (\pi\omega/2)} 
\]
\[
G_{++}(\omega) = G_{--}(\omega) = 
\]
\[
- \frac{4\sinh(g\omega/2N)}{\sinh(\pi\omega/2N)}\frac{\cosh[(\pi (N - 1) +
g)\omega/2N)}{\cosh (\pi\omega/2)}, 
\]
\begin{eqnarray}
G_{+-}(\omega,g) = G_{++}(\omega,\pi - g)  \label{kernels}
\end{eqnarray}
where ``+'' and ``-'' stand for particle and anti-particle and ``a,b'' label
phonon branches. 

In the limit $N \rightarrow \infty$ the system (\ref{TBA}) undergoes  a
drastic simplification. In particular, we get 
\begin{eqnarray}
G_{++}(\omega) = - g/\pi, G_{+-}(\omega) = - (1 - g/\pi)
\end{eqnarray}
that is the kernels in the TBA for the $\psi$-particle and antiparticle
become delta functions.

Another important fact is that in this limit the phonons decouple from the
TBA equations for ``+,--'' and can be analyzed separately. To pass to the $%
N\rightarrow \infty $ limit we introduce a continuous variable $y=a/N$ and
replace summation over $a,b$ by integration over $y$: 
\[
\sum_{a}\rightarrow N\int_{0}^{1}{\rm d}y
\]
Eqs.(\ref{TBA}) become 
\[
T\ln [1+{\rm e}^{\epsilon _{\pm }(\theta )/T}]-(1-g/\pi )T\ln [1+{\rm e}%
^{-\epsilon _{\pm }(\theta )/T}]
\]
\[
+(1-g/\pi )T\ln [1+{\rm e}^{-\epsilon _{\mp }(\theta )/T}]
\]
\begin{eqnarray}
&=&M\cosh \theta \pm \mu   \nonumber \\
&+&T\int_{0}^{1}{\rm d}y\int {\rm d}\theta ^{\prime }K_{+}(y,\theta -\theta
^{\prime })\ln [1+{\rm e}^{-\epsilon (y,\theta ^{\prime })/T}]  \label{part}
\end{eqnarray}
where the Fourier transform of the kernel is 
\[
K_{+}(y,\omega )=2g(1-g/\pi )\omega \sinh (y\pi \omega /2)/\cosh (\pi \omega
/2)
\]
and 
\begin{eqnarray}
&&T\ln [1+{\rm e}^{\epsilon (y,\theta )/T}]  \nonumber \\
&-&T\int_{0}^{1}{\rm d}y^{\prime }\int {\rm d}\theta ^{\prime }K(y,y^{\prime
};\theta -\theta ^{\prime })\ln [1+{\rm e}^{-\epsilon (y^{\prime },\theta
^{\prime })/T}]  \nonumber \\
&=&2M\sin (\pi y/2)\cosh \theta   \label{phonons}
\end{eqnarray}
where 
\[
K(y,y^{\prime };\omega )=4g(1-g/\pi )\omega \times 
\]
\[
\frac{\cosh [\pi (1-max(y,y^{\prime }))\omega /2]\sinh [\pi min(y,y^{\prime
})\omega /2]}{\cosh (\pi \omega /2)}
\]
Note  that terms with $\epsilon _{\pm }$ are absent in the last equation, as
expected in the large-$N$ limit where the  fermions  provide  a small
boundary effect on the phonons giving a $O(1/N)$-correction to their
thermodynamics.

Eq.(\ref{phonons}) is further simplified in the most interesting limit 
\begin{equation}
M/N \ll T \ll M
\end{equation}
where only the region $y \ll 1$ contributes to the thermodynamics. Defining
the new scaling variable $\xi = \pi y M/T$ we get 
\[
\ln[1 + {\rm e}^{\epsilon(\xi,\theta)/T}] + \gamma\frac{\partial^2}{%
\partial\theta^2}\int_0^{\infty}{\rm d} \xi^{\prime}min(\xi,\xi^{\prime})\ln[%
1 + {\rm e}^{- \epsilon(y^{\prime},\theta)/T}] 
\]
\begin{eqnarray}
= \xi\cosh\theta  \label{phonons1}
\end{eqnarray}
where $\gamma = 2g(\pi - g)(T/\pi M)^2$.

This equation together with the expression for the free energy 
\begin{eqnarray}
F_{phon}/NL = - \frac{T^3}{2\pi^2 M}\int_0^{\infty}{\rm d} \xi \xi{\rm d}%
\theta\cosh\theta \ln[1 + {\rm e}^{- \epsilon(\xi, \theta)/T}]
\label{freeph}
\end{eqnarray}
describes the continuous limit of the (2 + 1)-dimensional Toda array. As far
as we know such limit has never been described before. It follows from Eqs.(%
\ref{phonons1},\ref{freeph}) that temperature enters into $\epsilon/T$ only
through $\gamma$ and therefore 
\begin{eqnarray}
F_{phon}/NL = - \frac{T^3}{M}Y\left[2g(\pi -g )(T/\pi M)^2\right]
\end{eqnarray}
where $Y(\gamma)$ is a smooth positive function allowing regular expansion
in $\gamma$. This result is valid in  all powers of the bare coupling
constant $\beta$; being dependent on $g(\pi -g ) = 4\pi^3(4\pi/\beta +
\beta)^{-2}$ it is manifestly self-dual under transformation (\ref{dual}).

Eqs.(\ref{phonons1},\ref{freeph}) can be studied by iterations in $\gamma$.
For this purpose it is convenient to introduce the quantity 
\[
E(\xi,\theta) = \ln[1 + {\rm e}^{\epsilon(\xi,\theta)/T}] 
\]
such that these equations become 
\[
E(\xi,\theta) - \gamma\frac{\partial^2}{\partial\theta^2}\int_0^{\infty}{\rm %
d} \xi^{\prime}min(\xi,\xi^{\prime})\ln[1 - {\rm e}^{-
E(\xi^{\prime},\theta)}] = \xi\cosh\theta \label{phonons2} 
\]
\begin{eqnarray}
F_{phon}/NL = \frac{T^3}{2\pi^2 M}\int_0^{\infty}{\rm d} \xi \xi{\rm d}%
\theta\cosh\theta \ln[1 - {\rm e}^{- E(\xi, \theta)}]  \label{freeph2}
\end{eqnarray}
In this formulation it is absolutely clear that in the limit $g\rightarrow 0,\pi$
the Toda array becomes equivalent to the free (2+1)-dimensional bosonic
field and that the phonon free energy allows a regular expansion in $\gamma
\sim T^2$. Indeed, the substitution $p_x = M\xi\sinh\theta, p_y = M\xi$
brings Eq.(\ref{freeph2}) to the familiar form of the free energy of
two-dimensional acoustic phonons.  The first  terms in  the  expansion
in $\gamma$ are 
\begin{eqnarray}
Y(\gamma) = \frac{1}{\pi}\zeta(3) + \frac{3 \gamma}{2\pi^2}\int_0^{\infty}%
\frac{y dy}{e^y - 1}\int_0^y\frac{x^2 dx}{e^x - 1} + O(\gamma^2)
\end{eqnarray}

Further we shall be interested in the case when chemical potential overcomes
the gap $M$ and there is a finite density of charge in the ground state. In
this case the fermionic free energy is expanded in powers of $(T/\epsilon_F)$
where $\epsilon_F \approx \mu - M$ is the Fermi energy. 
In comparison with fermions
two-dimensional phonons give extra small factor $T/M$. Different temperature
dependences of the fermion and phonon sectors allow one to consider them
separately.

To illustrate this point let us consider T the last term in the right hand
side of Eq.(\ref{part}). At $T \ll M$ it becomes 
\begin{eqnarray}
\Sigma(\theta) &=& T\gamma \frac{\partial^2}{\partial\theta^2}%
G(\theta,\gamma) \sim T^3  \nonumber \\
G(\theta,\gamma) &=& \frac{\pi}{16}\int {\rm d} \xi \xi\int {\rm d}
\theta^{\prime}[\cosh(\theta - \theta^{\prime})]^{-1}\ln[1 - {\rm e}^{-
E(\xi,\theta^{\prime})}]  \label{self}
\end{eqnarray}
and can be treated as a thermal renormalization of the particles dispersion.
We shall omit it together with renormalization of the fermion free energy
coming from polarization of the host - the latter term being of order of $T^3
$ in comparison with $T^2$-contribution coming from the 1D particles.

After these simplifications the effective TBA for the fermions become

\begin{eqnarray}
\epsilon_+(\theta) + (g/\pi)T\ln[1 + {\rm e}^{-\epsilon_+(\theta)/T}] 
\nonumber \\
+ (1 - g/\pi)T\ln[1 + {\rm e}^{-\epsilon_-(\theta)/T}] = M\cosh\theta -\mu
\end{eqnarray}

\begin{eqnarray}
\epsilon_-(\theta) + (g/\pi)T\ln[1 + {\rm e}^{-\epsilon_-(\theta)/T}] 
\nonumber \\
+ (1 - g/\pi)T\ln[1 + {\rm e}^{-\epsilon_+(\theta)/T}] = M\cosh\theta + \mu
\end{eqnarray}
with the free energy equal to 
\begin{eqnarray}
F/L = - T\sum_{a =\pm} \frac{M}{2\pi}\int {\rm d}\theta \cosh\theta \ln[1 + 
{\rm e}^{- \epsilon_a(\theta)/T}]
\end{eqnarray}
These equations coincide with the general equations for particles with
exclusion statistics (\ref{excl}) for the particular choice of the matrix $%
G_{ab}$ given by (\ref{G}) (the fact that exclusion statistics exists in
this model has been first pointed out in \cite{Baseilhac}). As an
illustration we give their explicit solution for $g/\pi = 1/2$ and $T << M$: 
\begin{eqnarray}
F/L = - T\frac{M}{\pi}\int {\rm d}\theta \cosh\theta \ln\left[\frac{1}{2}%
{\rm e}^{- \epsilon_0(\theta)/T} + \sqrt{\frac{1}{4}{\rm e}^{-
2\epsilon_0(\theta)/T} + 1}\right]
\end{eqnarray}
where $\epsilon_0(\theta) = M\cosh\theta - \mu$.

In summary, we have found a relatively simple and realistic model  which
turns out to be a model of ideal gas of particles obeying  exclusion
statistics.  As a byproduct we have written down  equations for the free
energy of two-dimensional model of  anharmonic phonons. As we have pointed out in the introduction, one purpose of
this excersise is pedagogical. It has been pointed out many times
(for  instance, in \cite{Schoutens}, 
\cite{Krive}), that the ideal anyon gas in one dimension is just a
generalization of Luttinger liquid for non-linear
quasi-particle dispersion thus having  many properties common with Luttinger
liquids. On the
other hand,  analysis of such 
 physical processes as transport  looking rather
complicated in the standard bosonization approach is
greatly simplified in the anyon gas formalism (see \cite{Krive}). We
believe that the model discussed in this paper will be valuable for
such analysis. We also emphasise that the model discussed 
 is the first example of
a solvable model with  electron-phonon interaction. 

We are grateful to J. Cardy, P. Coleman, A. A. Nersesyan, 
I. Kogan, F. Essler, A.
Gogolin for  valuable remarks  and especially to V. A. Fateev for
hospitality and interest to the work.


\end{document}